\documentclass[twocolumn,prl,superscriptaddress,showpacs,showkeys,preprintnumbers,amsmath,amssymb]{revtex4}
\usepackage{dcolumn}
\usepackage{bm}
\usepackage[dvips]{graphicx}
\newcommand{\be}{\begin{equation}}
\newcommand{\ee}{\end{equation}}
\newcommand{\bea}{\begin{eqnarray}}
\newcommand{\eea}{\end{eqnarray}}

\begin{document}

\title{Alpha-Particle Clustering from Expanding Self-Conjugate Nuclei within the HFB Approach}

\author{M.~Girod}
\affiliation{CEA, DAM, DIF, F-91297 Arpajon Cedex, France}
\author{P.~Schuck}
\affiliation{Institut de Physique Nucl\'eaire, IN2P3-CNRS, Universit\'e Paris-Sud, F-91406 Orsay Cedex, France}
\affiliation{Laboratoire de Physique et Mod\`elisation des Milieux Condens\'es (LPMMC) (UMR 5493), Maison Jean Perrin, 
25 avenue des Martyrs BP 166, 38042 Grenoble cedex 9, France}

\date{Received: date/ Revised version: date}

\newcommand{\m}{\multicolumn}
\renewcommand{\arraystretch}{1.2}
\begin{abstract}
The nuclear equation of state is explored with the constrained HFB approach for self conjugate nuclei.
It is found that beyond a certain low, more or less universal density, those nuclei spontaneously
cluster into A/4 $\alpha$ particles with $A$ the nucleon number. The energy at
the threshold density increases linearly with the number of $\alpha$ particles as does the experimental
threshold energy. Taking off the spurious c.o.m. energy of each $\alpha$ particle almost gives agreement
between theory and experiment. The implications of these results with respect to $\alpha$ clustering
and the nuclear EOS at low density are discussed.
\end{abstract}
\pacs{21.60.Gx, 03.75.Fi, 21.10.Gv, 27.20.+n}
\maketitle

{\it Introduction.}
Cluster phenomena, in particular $\alpha$ particle clustering in lighter nuclei is presently a very active field of research.
 It is highlighted by the famous Hoyle ($0_2^+)$ state in $^{12}$C at 7.65 MeV.
This state, primordial for the $^{12}$C production in the universe and, thus, for life, 
is believed since long to be in good approximation formed out of a weakly interacting gas of almost 
free $\alpha$ particles \cite{Horiuchi}, \cite{Kamimura}, \cite{Uegaki}, \cite{THSR}.
Since these $\alpha$ particles are all in relative S-states, one can qualify this state 
as an $\alpha$ particle condensate \cite{THSR}
 keeping in mind the limitations of this notion for finite systems with small numbers of particles.
The research concerning this state has known a very vivid revival since about ten years when the hypothesis
of the possible existence of $\alpha$ condensates in nuclei was formulated for the first time \cite{THSR}.
The investigations are now extending to heavier self-conjugate nuclei. On the forefront is $^{16}$O where
theoretical investigations predict that the 6-th $0^+$ state at 15.1 MeV is an analogue of the Hoyle
state but with four $\alpha$ particles instead of three \cite{Funaki}. Similarities between the three $\alpha$ and
the four $\alpha$ cases are, indeed, being found experimentally \cite{Itoh}. The particularity of those
 $\alpha$ particle condensate states is that they are spatially extended \cite{a-rev}, i.e., at a low average density
 of $\rho \sim \rho_{eq} /3-\rho_{eq} /4$ with $\rho_{eq}$ the average density at equilibrium of the nucleus. 
In this sense the $\alpha$ condensate states can be considered as a continuation of the structure of $^8$Be which consists of
two well identifiable, separated, weakly interacting $\alpha$ particles with average density
 in the just mentioned range \cite{Wiringa}.
On the other hand it is also well known that low density nuclear matter is unstable against cluster formation,
 mainly $\alpha$ particles \cite{Ropke}, \cite{Takemoto}. Theoretical predictions give a critical temperature
 for macroscopic $\alpha$ condensation as high as $T_c^{\alpha} \sim 7-8$ MeV at low densities \cite{Sogo}. 
From this fact, it can be inferred that the Hoyle state and possible heavier Hoyle analogue states are precursor
states of a macroscopic $\alpha$ condensate phase, very much in analogy to neutron pairing in finite nuclei being
a precursor to neutron superfluidity in neutron stars.

The description of $\alpha$ gas states in heavier n$\alpha$ nuclei naturally becomes more and more difficult 
using, e.g., $\alpha$ condensate wave functions as they are given by the THSR wave function \cite{Tohsaki}
 which is based on a fully fermionic description. On the other hand,
certain 3D Hartree-Fock (HF) and Hartree-Fock-Bogoliubov (HFB) calculations of nuclei have recently shown that
these mean field approaches can manifest  cluster formation  \cite{GirodM}, \cite{Maruhn}, \cite{Vretenar}. 
They are less affected by size limitations.
In this work, we concentrate within the HFB framework, using the Gogny D1S interaction, on constraining the radius
of self-conjugate nuclei to larger and larger values, i.e., to lower and lower nuclear densities. In this way, we prevent
a transition to strong deformation which would favor clusterization into binaries. Thus, expanding the nucleus,
at a critical low density and because of the 3D nature of the code, the system will spontaneously cluster
into $\alpha$ particles, eventually also into a heavier compact core with an $\alpha$ gas around it and
other cluster formations.   Those $\alpha$ particles do, of course, not form a condensate but rather build
a lattice. This hinges on the fact that the $\alpha$'s have not the possibility to move freely with their
center of mass (c.o.m.) coordinate in these HF or HFB calculations. The advantage of the mean field approach
is that it can produce many $\alpha$'s in various configurations, still being entirely microscopic.
So, qualitatively, the transition of an expanding nucleus passing from the homogeneous density distribution
of a Fermi gas (HF) to clusterization can be studied within the mean field approach giving precious insights
into the clusterization phenomenon in general and into the formation of  $\alpha$ gas phases in particular.
For example, as we will show,  the energy of the system as a function of the radius first raises from its
equilibrium position going over a barrier and entering the cluster phase at around a density $\rho = \rho_{eq}/3$.
Among others, this feature is of quite some interest as will be discussed below. 

\noindent
{\it Formalism and Results.}
Since the constrained HFB theory is extensively explained in the literature~\cite{girod,triax,RS,gogny},
we here only give the absolute minimum of formalism.
We minimize the HFB ground state energy using the Gogny D1S~\cite{gogny} interaction
in constraining the radius of the nucleus, that is
\begin{equation}
E^{\mbox{HFB}}=\langle\mbox{HFB} |H-\lambda r^2|\mbox{HFB}\rangle/\langle \mbox{HFB}|\mbox{HFB}\rangle ,
\end{equation}
\noindent
where $r$ is the radius.  $\lambda$ is obtained in such way that $\langle\mbox{HFB} |r^2|\mbox{HFB}\rangle = r_0^2$.
Therefore, choosing  values for $r_0 <$ or $> r_{GS}$, where $r_{GS}$ is the radius of the ground state, 
compresses or dilutes the nucleus.
In the forthcoming, we treat all nuclei in spherical geometry, even though HFB may sometimes yield a deformed solution at
the equilibrium position. Since we are interested in the low density (large radius) regime, it does not matter what is
precisely the configuration at the absolute minimum. It should, however, be stressed that our 3D numerical code allows
to take on any cluster configuration, if this is energetically favorable but on average the system stays spherical.
 For our study, we consider selfconjugate N=Z nuclei up to  $^{40}$Ca \cite{corner}.

\begin{figure}
\begin{center}
\includegraphics[width=7cm]{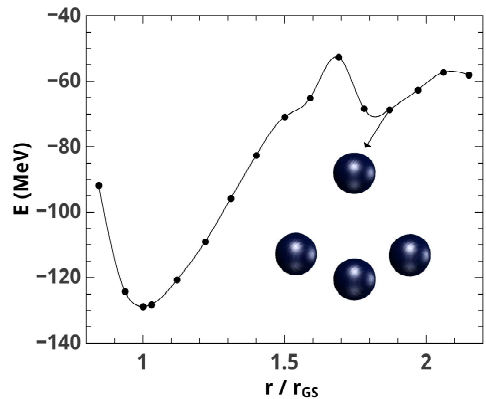}
\caption{(Color online) Total energy of $^{16}$O as a function of the radius scaled
with respect to the one of the ground state $r_{GS}$. At $r / r_{GS}$ = $\sim 1.8$, we see that a tetrahedron
of four $\alpha$ particles is formed. No c.o.m. correction for individual $\alpha$'s is applied here. The arrow indicates to which $r/r_{GS}$ value the $\alpha$ configuration corresponds to. }
\label{fig3}
\end{center}
\end{figure}

\begin{figure}
\begin{center}
\includegraphics[width=7cm]{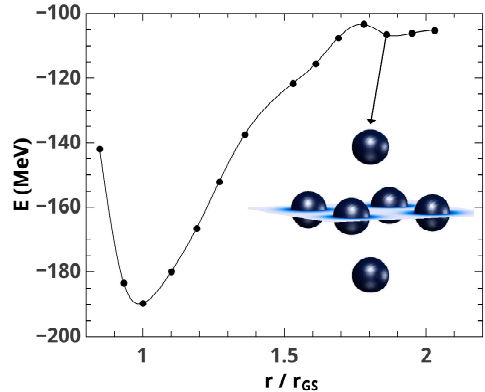}
\caption{(Color online) Same as Fig.1 but for $^{24}$Mg with six $\alpha$'s. The shaded area only serves to show the three dimensionality of the $\alpha$ arrangement.}
\label{fig5}
\end{center}
\end{figure}

\begin{figure}
\begin{center}
\includegraphics[width=7cm]{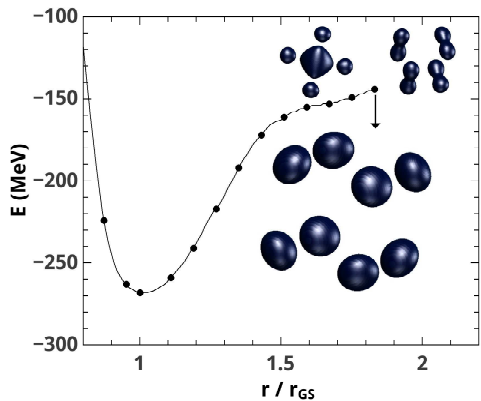}
\caption{(Color online) Same as Fig.1 but for $^{32}$S with eight $\alpha$'s.
Also configurations with four $^8$Be's and a $^{16}$O surrounded by four $\alpha$'s are shown.}
\label{fig6}
\end{center}
\end{figure}

\begin{figure}
\begin{center}
\includegraphics[width=7cm]{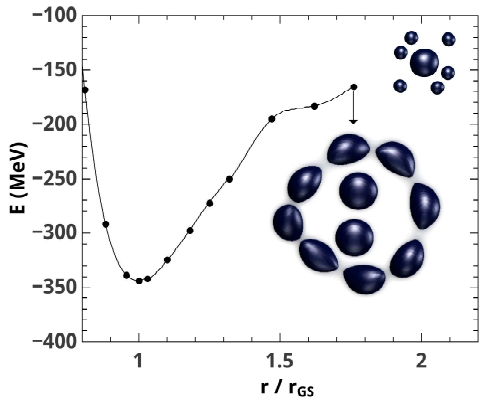}
\caption{(Color online) Same as Fig.1 but for $^{40}$Ca with ten $\alpha$'s. Also configurations with
 a $^{16}$O surrounded by six $\alpha$'s is shown}
\label{fig7}
\end{center}
\end{figure}

\noindent
Let us first show in some detail the various $\alpha$ cluster configurations obtained from our constrained HFB calculation
 (for space reason, we will not show in this work the well known triangle configuration of $^{12}$C, see, e.g.,
  \cite{Neff,Gulmi}).
In Fig.~\ref{fig3}, we present the $^{16}$O case. We see that a tetrahedron of four $\alpha$ particles
is formed. Actually the transition to the cluster state is quite abrupt. 
In Fig.~\ref{fig5} we show the $^{24}$Mg case. The $^{20}$Ne case is quite similar, 
only in the shaded plane three $\alpha$'s are arranged in an equilateral triangle instead of four at the corners of a square. 
In Fig.~\ref{fig6} we display $^{32}$S and in Fig.~\ref{fig7} $^{40}$Ca.
Going to the heavier systems, it becomes more
and more difficult to disrupt the system into $\alpha$ particles only. For example we show a four $^8$Be configuration
for $^{32}$S and a $^{16}$O plus six $\alpha$ case for $^{40}$Ca. Many more cluster configurations 
can be obtained progressing,
e.g. in smaller steps with the radius increment but for space reasons we cannot present this here. Let us also mention 
that we got an excited $^{36}$Ar consisting out of three $^{12}$C in a bent linear chain configuration. 
Also $^{48}$Cr clustering into four $^{12}$C has been found, and many configurations more.

\begin{figure}
\begin{center}
\includegraphics[width=8.6cm]{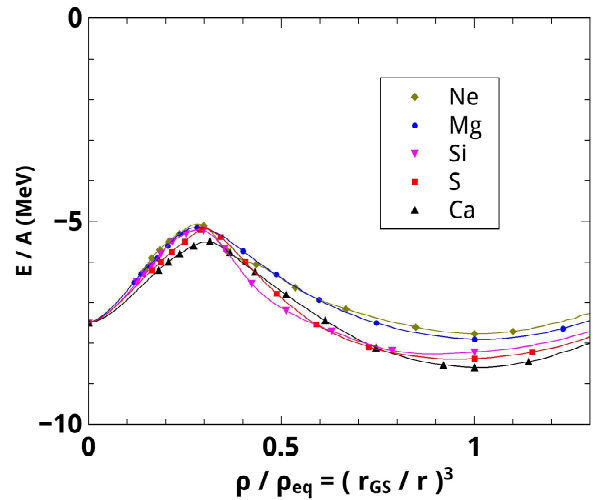}
\caption{(Color online) Equation of state for a choice of self-conjugate nuclei (EOS-A) as a function of average density 
scaled by the one at equilibrium,
see text for detailed definition.}
\label{fig1}
\end{center}
\end{figure}

Let us now present the equation of state for the energy per particle as a function of density.
Expanding (or compressing) a finite spherical nucleus yields, of course, not the usual equation 
of state as in infinite nuclear matter,
since besides the bulk also surface and Coulomb energies together with the quantal shell corrections are involved.
Therefore, this equation of state which we want to call EOS-A slightly differs from nucleus to nucleus.
Even for a given nucleus, in the low density region where clusters are formed, EOS-A may fluctuate, since in this region
the energy surface has many different valleys leading to different cluster formations not very much different in energy.
In which configuration the calculation gets trapped depends, e.g., on the step size of the expansion and other ingredients. 
It is important to realise that, once the $\alpha$ particles are formed, 
in HFB they contain their own spurious c.o.m. energy which should be eliminated. 
Since presently no method is available to achieve this in a microscopic way, we follow a heuristic procedure.
We perform an HFB calculation of ${^8}$Be and constrain the distance between the two
forming $\alpha$'s so that they are very well separated. About 14 MeV are then missing to get twice the binding energy of
a single $\alpha$ particle in the asymptotic limit, as it should be.
We attribute this lack of binding to spurious c.o.m. motion of each $\alpha$ not being correctly treated.
Of course, the total kinetic energy is subtracted from the Hamiltonian in all our calculations.
So for ${^8}$Be we have $\sim$ 7 MeV extra binding per $\alpha$ particle. We make the hypothesis that this number
stays about the same, even in cases with more $\alpha$ particles. 
This correction to take off 7 MeV for each $\alpha$ particle is switched on adiabatically from the point 
of the first clear appearance of the $\alpha$ particle structure what happens around 
a density $\rho/\rho_{eq.} \sim 1/3$.
In order to get a global view, we show in Fig.~\ref{fig1} the different  EOS-A 
obtained in this way for various n-$\alpha$ nuclei superposed.
With this, we want to put into evidence the general behavior of the nuclear equation of state
at low densities when it goes over into an $\alpha$ particle configuration. As can be seen from Fig.~\ref{fig1},
there is a clear tendency that the EOS-A goes as a function of decreasing density over a maximum before
reaching the zero density limit where the $\alpha$ particles are infinitely far apart and, therefore,
the EOS-A reaches the value of an isolated $\alpha$ particle, i.e. -7.5 MeV which is our theoretical value.
Evidently the numerical HFB code
cannot handle configurations with $\alpha$ particles very distant from one another. Therefore, we stopped the calculation,
once the $\alpha$ particles are clearly separated what happens around $\rho \sim \rho_{eq}/5$
(see also the detailed figures above). It may seem intriguing that the EOS-A bends down at low densities 
even for $^{32}$S and $^{40}$Ca where the energies displayed in Figs. 3 and 4 still show a slight increase in energy.
 It should, however, be recalled that the energies shown in Figs. 1-4 are uncorrected for spurious $\alpha$ 
particle c.o.m. motion. Once this correction is applied, the slight upward trend is converted into a downward trend.
In Fig.~\ref{fig1}, we show as an artist view lines extrapolating down to zero
density just to guide the eye.
The existence of a maximum in the nuclear equation of state containing a gas of $\alpha$ particles on the low density side
and a Fermi gas (HF) on the higher density side is not evident.
 It would mean that the $\alpha$ phase is in a meta-stable state.
The transition to the Fermi gas configuration will be strongly different from the scenario when there is no barrier.
This may be a question eventually of importance in compact stars where $\alpha$ particle phases may exist
in the density-temperature space. 
This question has been investigated in recent years by several authors, see \cite{Toki,Schwenk,Roepke} 
but the existence of a barrier and its hight has been discussed, to the best of our knowledge, 
only in a relatively older paper
 on nuclear matter by Takemoto et al \cite{Takemoto}  with similar results to ours.
The present investigation seems to indicate the existence
of a barrier about 2.5 MeV high but certainly our procedure is very crude and 
more investigations have to be performed before a definite conclusion can be made. 
It should, however, be observed that at $\rho/\rho_{eq} \sim 1/3$ where the $\alpha$'s start to appear, 
the EOS-A are already well above the asymptotic limit of -7.5 MeV, so that in any case the systems have 
to go over a substantial barrier. This is the important point.
The existence and hight of the barrier are, of course, of great importance for the coalescence 
process of $\alpha$ particles into heavier nuclei in such star scenarios.

\begin{figure}
\begin{center}
\includegraphics[width=8.6cm]{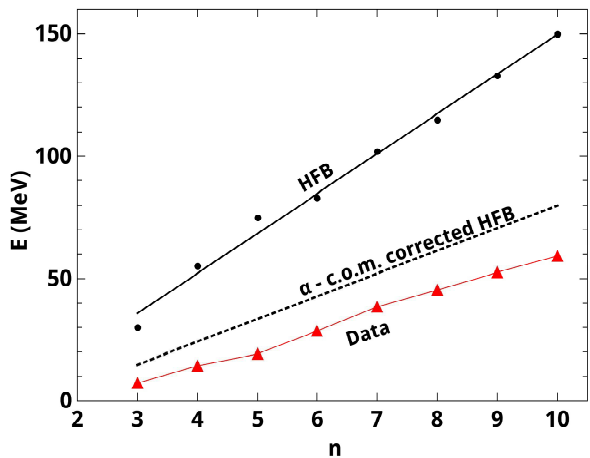}
\caption{(Color online) Threshold energies as a function of the number
$n$ of $\alpha$ particles. Triangles: experimental values; dots: values from HFB calculations,
see text for precise definition; full line: best straight line fit to HFB results; broken line: alpha particle c.o.m. corrected HFB values.}
\label{fig2}
\end{center}
\end{figure}

 Defining $\rho= \rho_{eq}/3~ (r/r_{GS} \sim 1.45)$ as the theoretical threshold for $\alpha$ formation,
we display in Fig.~\ref{fig2} the energy progression with the number n of $\alpha$ particles at that density.
It is seen that this progression is about linear with n, increasing by $\sim 16 $ MeV per $\alpha$ particle.
Taking off 7 MeV of spurious c.o.m. energy for each $\alpha$ particle strongly improves the agreement with experiment,
see the broken line in Fig.~\ref{fig2}.
The experimental threshold energies follow rather well a 7.6 MeV increase per $\alpha$ particle.
It is, however, clear that this procedure can only yield a very rough estimate of the real situation.
It is encouraging that the overall picture seems to be quite reasonable.
Since it is clear by now that
the $\alpha$ particles form a quantum gas rather than a crystal, see \cite{Bo} where a Brink-type,
i.e. crystal-like of approach is put into competition with the THSR approach with the latter the clear winner,
it will be important for the future to find less heuristic ways to take off the spurious c.o.m. energies
from the clusters, once they are formed in the mean field approach.

\noindent
{\it Summary and Discussion.}
In this work, for the first time, a rather systematic study for quite a number of self-conjugate
 nuclei is presented within mean field theory (HFB) concerning the formation of $\alpha$ particles 
when the nuclei are expanding that is,
at low density. We here adopted a static approach revealing rich scenarios of $\alpha$ cluster configurations
and other heavier clusters like $^8$Be and $^{12}$C. However, for the lighter nuclei $\alpha$ clusters are largely dominant.
The mean field approach has the great advantage over other cluster models to be entirely microscopic employing a
realistic energy density functional and to be able to describe the formation of quite a large number of $\alpha$
particles and eventually other clusters. It can cover within the same approach all density regions going in a
continuous way from stable nuclei to highly excited ones at low density where the clusters form. It is found in
this work that expanding an n-$\alpha$ nucleus the corresponding EOS-A goes over a maximum before reaching
 the asymptotic very low density limit of the $\alpha$ gas. This may be of importance in stabilizing an $\alpha$ phase.
In principle there is no restriction for our 3D mean field approach to produce any kind of shapes and clusters 
in which the systems want to go into.
We also have checked that a single $\alpha$ particle is well described in HF with the Gogny force.
Indeed, we have demonstrated in this work that there can exist a great variety of rather
surprising and unexpected cluster configurations when the nucleus is expanding.

The disadvantage
of the mean field approach is that it fixes the clusters to certain spatial positions as, e.g., on the corners 
of a tetrahedron in the case of $^{16}$O whereas it is predicted
in recent work with the so-called THSR wave function that $\alpha$ particles rather form a (degenerate quantum)
gas than a crystal \cite{Horiuchi,Kamimura,Uegaki,THSR,Yamada}. 
To overcome this drawback, we applied in this work a purely heuristic procedure in
eliminating 'by hand' the spurious c.o.m. energy of each $\alpha$ particle. It is shown that in this way the
theoretical threshold energies for n $\alpha$'s get rather close to the experimental values, see Fig. 2. 
Let us mention that other approaches have also been used before for the description of cluster formation \cite{Neff,Gulmi}. 
This was mostly done within the Antisymmetrised Molecular Dynamics (AMD) \cite{Enyo} or 
Fermionic Molecular Dynamics (FMD) \cite{Neff} approaches.
 We are not aware of any work which uses HF or HFB with wave functions in a systematic study for
 clustering at low densities.
The correct microscopic treatment of the spurious c.o.m. motion of clusters formed in a mean field approach remains
an important task of nuclear many body physics for the future.
Our work opens a variety of further investigations. Most interesting is the cluster formation
as a function of neutron excess.
Repeating our study with  Relativistic Mean Field (RMF) may also be interesting, since it recently has been shown that
RMF favors cluster formation \cite{Vretenar}.
We believe that the rich cluster scenarios found in this work are very inspiring and we hope that this will trigger
more experimental and theoretical work on this line in the future.\\

\noindent
{\it Acknowledgments.}
P.S. is very grateful to Y. Funaki, H. Horiuchi, G. R\"opke, A. Tohsaki, and T. Yamada for a very fruitful and longstanding
 collaboration on $\alpha$ clustering in nuclear systems where many ideas developed which are alluded to in this work.
More recent collaboration and discussions with Bo Zhou, Zhongzhou Ren, and Chang Xu also have been very helpful.
Exchanges of ideas with J.-P. Ebran, E. Khan, D. Vretenar are appreciated.


\begin{thebibliography}{5}
\bibitem{Horiuchi} H. Horiuchi, Prog. Theor. Phys. {\bf51}, 1266 (1974); {\bf53}, 447 (1975).
\bibitem{Kamimura} M. Kamimura, Nucl. Phys. {\bf A351}, 456 (1981).
\bibitem{Uegaki} E. Uegaki, S. Okabe, Y. Abe, H. Tanaka, Prog. Theor. Phys. {\bf57}, 1262 (1977); {\bf59}, 1031 (1978); {\bf62}, 1621 (1979).
\bibitem{THSR} A. Tohsaki, H. Horiuchi, P. Schuck, G. R\"opke, Phys. Rev. Lett. {\bf87}, 192501 (2001).
\bibitem{Funaki} Y. Funaki, T. Yamada, H. Horiuchi, G. R\"opke, P. Schuck, A. Tohsaki, Phys. Rev. Lett. {\bf101}, 082502 (2008).
\bibitem{Itoh} M. Itoh et al., talk given at 10th Int. Conf. on Clustering Aspects of Nuclear Structure and Dynamics, Debrecen, Hungary, 2012, J. Phys.: Conf. Ser. {\bf436}, 012006 (2013).
\bibitem{a-rev} Y. Funaki, A. Tohsaki, H. Horiuchi, P. Schuck, G. Roepke, Eur. Phys. J. {\bf A28}, 259 (2006).
\bibitem{Wiringa} R.B. Wiringa, S.C. Pieper, J. Carlson, V.R. Pandharipande, Phys. Rev. C {\bf62}, 014001 (2000).
\bibitem{Ropke} G. R\"opke, L. M\"unchow, H. Schulz, Nucl. Phys. {\bf A379}, 536 (1982); G. R\"opke, M. Schmidt, L. M\"unchow, H. Schulz, Nucl. Phys. {\bf A399}, 587 (1983).
\bibitem{Takemoto} H. Takemoto, M. Fukushima, S. Chiba, H. Horiuchi, Y. Akaishi, A. Tohsaki, Phys. Rev. C {\bf69}, 035802 (2004).
\bibitem{Sogo} T. Sogo, R. Lazauskas, G. R\"opke, P. Schuck, Phys. Rev. C {\bf79}, 051301 (2009).
\bibitem{Tohsaki} A. Tohsaki, Frontiers of Physics Vol. {\bf6}, 320 (2011).
\bibitem{GirodM} Y. Funaki, M. Girod, H. Horiuchi, G. R\"opke, P. Schuck, A. Tohsaki, T. Yamada, J. Phys. G {\bf37}, 064012 (2010).
\bibitem{Maruhn} T.Ichikawa, J.A. Maruhn, N. Itagaki, S. Ohkubo, Phys. Rev. Lett. {\bf107}, 112501 (2011); T.Ichikawa, J.A. Maruhn, N. Itagaki, K. Matsuyanagi, P.-G. Reinhard, S. Ohkubo, Phys. Rev. Lett. {\bf109}, 232503 (2012).
\bibitem{Vretenar} J.-P. Ebran, E. Khan, T. Niksic, D. Vretenar, Phys. Rev. C {\bf87}, 044307 (2013);
J.-P. Ebran, E. Khan, T. Niksic, D. Vretenar, Nature {\bf487}, 341 (2012).
\bibitem{girod} M. Girod, and B. Grammaticos, Phys. Rev. C {\bf27}, 2317 (1983).
\bibitem{triax}  M. Girod, J.P. Delaroche, D. Gogny, J.F. Berger, Phys. Rev. Lett. {\bf62}, 2452 (1989).
\bibitem{RS} P. Ring, P. Schuck, The Nuclear many Body  Problem, Springer 1980.
\bibitem{gogny} J. Decharg\' e and D. Gogny, Phys. Rev.  C {\bf 21}, 1568 (1980); J.F. Berger, M. Girod, and D. Gogny, Comp. Phys.Comm. {\bf 63}, 365 (1991).
\bibitem{corner} In principle, for very strong contraints, it could happen that the expansion process becomes unstable and all nucleons migrate to the corners. We are here still far from this limit. The systems are also stabilized due to the use of harmonic oscillator basis functions.
\bibitem{Neff} M. Chernykh, H. Feldmeier, T. Neff, P. von Neumann-Cosel, A. Richter, Phys. Rev. Lett. {\bf98}, 032501 (2007).
\bibitem{Gulmi} T. Furuta, K.H.O. Hasnaoui, F. Gulminelli,, C. Leclercq, A. Ono, Phys. Rev. C {\bf 82}, 034307 (2010).
\bibitem{Toki} H. Shen, H. Toki, K. Oyamatsu, K. Sumiyoshi, Prog. Theor. Phys. {\bf100}, 1013 (1998).
\bibitem{Schwenk} C.J. Horowitz, A. Schwenk, Nucl. Phys. {\bf 776}, 55 (2006)
\bibitem{Roepke} G. Roepke, arXiv: 0810.4645.
\bibitem{Bo} Bo Zhou, Y. Funaki, H. Horiuchi, Zhongzhou Ren, G. R\"opke, P. Schuck, A. Tohsaki, Chang Xu, T. Yamada,
 Phys. Rev. Lett. {\bf110}, 262501 (2013).
\bibitem{Yamada} T. Yamada, Y. Funaki, H. Horiuchi, G. Roepke, P. Schuck, A. Tohsaki, `` Cluster in Nuclei (Lecture Notes in Physics)-Vol.2''. ed. by C. Beck, (Springer-Verlag, Berlin, 2011).
\bibitem{Enyo} Y. Kanada-En'yo, M. Kimura, A. Ono, arXiv: 1202.1864.

\end{thebibliography}
\end{document}